\documentclass[12pt,a4paper]{article}
\usepackage{amsmath,amsfonts}
\usepackage{graphicx}

\begin{document}
\begin{flushright}
%hep-th/yymmnnn\\
NA-DSF-08/2007
\end{flushright}
\vfill
\thispagestyle{empty}

\renewcommand{\thefootnote}{\fnsymbol{footnote}}

\begin{center}
\textbf{\Large Stretched Horizon and Entropy of Superstars}\\[48pt]%
Luca D'Errico, Wolfgang M\"uck and Roberto Pettorino\footnote[1]{E-mail addresses: \texttt{mueck,pettorino@na.infn.it}}\\[24pt]%
\textit{Dipartimento di Scienze Fisiche, Universit\`a degli Studi di
  Napoli ``Federico II''}\\%
\textit{and INFN, Sezione di Napoli, via Cintia, 80126 Napoli, Italy}%
\end{center}
\vfill

\begin{abstract}
Amongst the class of supergravity solutions found by Lin, Lunin and Maldacena, we consider pure and mixed state configurations generated by phase space densities in the dual fermionic picture. A one-to-one map is constructed between the phase space densities and piecewise monotonic curves, which generalize the Young diagrams corresponding to pure states. Within the fermionic phase space picture, a microscopic formula for the entropy of mixed states is proposed. Considering thermal ensembles, agreement is found between the thermodynamic and the proposed microscopic entropies. Furthermore, we study fluctuations in thermodynamic ensembles for the superstar and compare the entropy of these ensembles with the area of stretched horizons predicted by the mean fluctuation size. 
\end{abstract}
\vfill \vfill \vfill
\newpage
% end preamble

\renewcommand{\thefootnote}{\arabic{footnote}}

% definitions

\newcommand{\ie}{\emph{i.e.},}
\newcommand{\eg}{\emph{e.g.},}

\newcommand{\rmd}{\mathrm{d}}

\newcommand{\e}[1]{\operatorname{e}^{#1}}

\newcommand{\bx}{\mathbf{x}}

\newcommand{\tOmega}{\tilde{\Omega}}

\newcommand{\vev}[1]{\left\langle #1 \right\rangle}

\newcommand{\Var}{\operatorname{Var}}

\newcommand{\rmLi}{\operatorname{Li}}

\newcommand{\tr}{\operatorname{tr}}

\newcommand{\Order}[1]{\mathcal{O}(#1)}

\section{Introduction}
\label{intro}

Extremal small black holes are intriguing objects. On the one hand, as solutions of supergravity (SUGRA), they have vanishing horizon area, because the horizon locus coincides with the space-time singularity. On the other hand, counting microstates yields a finite entropy (for a review, see \cite{Mathur:2005ai}). Within String Theory, the origin of this discrepancy is to be found in the supergravity approximation. In simple terms, in the supergravity limit, one considers objects of the size comparable to a typical gravitational length scale, $L$, which is much larger than the string length, $L \gg {\alpha'}^{1/2}$. Consequently, objects of a typical size $\lambda\ll L$ cannot appear, even if $\lambda \gg {\alpha'}^{1/2}$. This is the case of small black holes. The horizon area that would follow from their entropy and the Bekenstein-Hawking formula, $A=4G S$, vanishes in the supergravity limit when measured in units of $L$. Including higher derivative corrections to SUGRA generates intermediate length scales; \eg\ one could have $\lambda^4 = \alpha' L^2$. Therefore, for small black holes, a finite-size horizon is expected to be generated, and the coefficient $1/4$ in the Bekenstein-Hawking formula can receive large corrections \cite{Dabholkar:2004dq}.

Maybe the easiest way conceptually to find the entropy of extremal small black holes is to use Sen's entropy function \cite{Sen:2005wa} for a SUGRA Lagrangian with higher derivative corrections \cite{Morales:2006gm, Sinha:2006yy}.\footnote{A similar method is $c$-function extremization \cite{Kraus:2005vz}. In this method,
higher derivative corrections have been considered in \cite{Castro:2007sd}. A generalization of Sen's entropy function formalism for rotating black holes has been developed in \cite{Astefanesei:2006dd}.} 
%Unfortunately, higher derivative corrections are not yet available for all SUGRA theories of interest.
Another way, which yields, however, only an order-of-magnitude estimate of the entropy, employs the fuzzball conjecture (for a review, see \cite{Mathur:2005zp}). According to this conjecture, the black hole geometry is the result of coarse graining the underlying microstates, which possess distinct geometries differing from each other only in the space-time region around the singularity. To observers far from the black hole measuring typical observables (in the sense discussed in \cite{Balasubramanian:2005mg}), the differences between microstates are invisible. Placing a stretched horizon around the space-time region, where the microstates significantly differ, one can estimate the black hole entropy using the Bekenstein-Hawking area law. 

In this article, we consider the bubbling solutions found by Lin, Lunin and Maldacena (LLM) \cite{Lin:2004nb},   which are the gravity duals of $1/2$-BPS chiral primary operators in $\mathcal{N}=4$ Super Yang-Mills (SYM) theory. The $1/2$-BPS sector has a simple description in terms of free fermions in a harmonic oscillator potential, and an orthogonal set of states is in one-to-one correspondence with Young diagrams \cite{Corley:2001zk, Berenstein:2004kk}. The LLM bubbling solutions have been extensively studied in the literature. Duality between the Pauli exclusion principle and the absence of closed time-like curves has been established in \cite{Caldarelli:2004mz, Milanesi:2005tp}. Statistical ensembles of bubbling solutions have been considered in 
\cite{Buchel:2004mc, Suryanarayana:2004ig, Silva:2005fa,  Balasubramanian:2005mg}. Further work can be found in \cite{Ghodsi:2005ks, Takayama:2005bc, Takayama:2005yq, Maoz:2005nk, Giombi:2005zq, Liu:2006pd, Sato:2007zu, Chen:2007gh, Brown:2007bb}. 

In \cite{Balasubramanian:2005mg}, an ensemble of microstates has been proposed that fits the geometry of the superstar \cite{Behrndt:1998ns, Behrndt:1998jd, Cvetic:1999xp}. In the present article, we will further test this ensemble by calculating an entropy estimate using a stretched horizon, lifting the geometric analysis of \cite{Suryanarayana:2004ig} to ten dimensions. To this end, we shall consider the mean size of fluctuations of the superstar ensemble, which provides a measure for the size of the stretched horizon, with the result that the predicted horizon size is too large in orders of magnitude. Hence, we will propose another ensemble, which we shall call the \emph{restricted superstar ensemble}, which also gives rise to the superstar geometry, but whose fluctuation sizes are consistent with the order of magnitude of the superstar entropy. On the way, we will discuss in detail the map between Young diagrams that characterize microstates and fermion droplets that generate the LLM geometry, as well as the generalization of that map to mixed states. Moreover, we will propose and test a unique formula for the entropy of mixed states, which arises directly from identifying the LLM plane of droplets with the semi-classical phase space of fermions. 

The article is organized as follows. In Sec.~\ref{review}, we briefly review the LLM solutions and some of their properties. Sec.~\ref{young} deals with the one-to-one map between fermion phase space density functions $u(r)$ and the (piecewise) monotonic curves $Y(X)$, which generalize the Young diagrams associated with microstates. At the end of that section, we propose our new entropy formula. In Sec.~\ref{superstar}, we discuss the construction of a stretched horizon in the superstar geometry, lifing the 5d analysis of \cite{Suryanarayana:2004ig} to ten dimensions. This will allow us to make contact later with the typical size of fluctuations in ensembles of Young diagrams. In Sec.~\ref{ensembles}, some statistical ensembles of Young diagrams are considered. We shall start with a review of the general grand canonical ensemble of \cite{Balasubramanian:2005mg}, by means of which we will be able to test our entropy formula. Then, we shall specialize to the grand canonical superstar ensemble of \cite{Balasubramanian:2005mg} and find that the size of the stretched horizon compatible with the fluctuations of that ensemble is too large. Therefore, we define and study a restricted superstar ensemble, whose typical fluctuation size matches the expected size of a stretched horizon. Finally, Sec.~\ref{conclusions} will contain conclusions.

\section{Review of LLM solution}
\label{review}

The metric of the LLM solutions is given by
\begin{equation}
\label{review:metric}
  \rmd s^2 = -h^{-2} (\rmd t +V_i \rmd x^i)^2
           +h^2 (\rmd y^2 +\rmd x^i \rmd x^i) 
           +y\e{G} \rmd \Omega_3^2
           +y\e{-G} \rmd \tOmega_3^2~,
\end{equation}
with $i=1,2$, and 
\begin{equation}
\label{review:hdef}
  h^{-2} = \frac{2y}{\sqrt{1-4z^2}}~, \qquad
  \e{2G} = \frac{1+2z}{1-2z}~.
\end{equation}
The auxiliary function $z$ and the vector $V_i$ are given by
\begin{equation}
\label{review:zVsols}
\begin{split}
  z(\bx,y) &= \frac{y^2}{\pi} \int \rmd^2 x'\, 
       \frac{z_0(\bx')}{[(\bx-\bx')^2 +y^2]^2}~, \\  
  V_i(\bx,y) &= \frac{\epsilon_{ij}}{\pi} \int \rmd^2 x'\, 
       \frac{z_0(\bx')(x-x')^j}{[(\bx-\bx')^2 +y^2]^2}~,
\end{split}
\end{equation}  
where $z_0(\bx) = z(\bx,0)$ is the boundary data for the auxiliary function $z(\bx,y)$ on the
plane $y=0$, spanned by $x^1$ and $x^2$. For the complete solution including the 5-form, we refer the
reader to the original paper \cite{Lin:2004nb}.

The metric \eqref{review:metric} is regular, if
$z_0(\bx)=\pm 1/2$. It possesses naked null-singularities for $|z_0|<1/2$,
and closed time-like curves (CTCs) for $|z_0|>1/2$
\cite{Caldarelli:2004mz}.\footnote{Clearly, one must cut out those regions of the
  $(\bx,y)$ 3-space, where $|z(\bx,y)|>1/2$. Then, the metric is
  singular at the space-time boundary, where $|z(\bx,y)|=1/2$ for 
  $y>0$. CTCs are found within space-time, close to
  this boundary.}
In this paper, we shall consider only solutions with $|z_0|\leq 1/2$.

It is convenient to define 
\begin{equation}
\label{review:udef}
  z_0(\bx) = \frac12 - u(\bx)~,
\end{equation}
so that, for a regular metric, only $u(\bx)=0,1$ are allowed values. 
For better visualization, one can think of $u$ as specifying
a certain darkness, or grayscale, on the 1-2-plane, 
with $u=0$ and $u=1$ corresponding to colours white and black, 
respectively. Then, the regular solutions are
characterized by ``droplets'', \ie\ arbitrary configurations of black
and white regions on the 1-2-plane, whereas gray areas 
generate naked null-singularities. 

Intuitively, one can identify $u(\bx)$ with the semi-classical phase-space
density of fermions in the dual fermionic picture that describes the
1/2-BPS sector. In this context, the Pauli exclusion principle for fermions (and
holes), $0\leq u \leq 1$, implies the absence of CTCs \cite{Caldarelli:2004mz}. The regular
solutions described by a distribution of black droplets
are the duals of pure states, or microstates, whereas
solutions that involve gray areas are duals of mixed states.

In what follows, we shall consider solutions corresponding to a large,
but finite, number of fermions, $N$. Identifing the 1-2-plane with the
semi-classical phase space of the fermions, we should remember that it is
``discretized'' into small elements of volume $2\pi\hbar$. For small
$\hbar$, we approximate the sum over these phase space elements as 
\begin{equation}
\label{review:phspace.sum}
  \sum\limits_{\text{ph.sp.}} = \frac{1}{2\pi\hbar} \int \rmd^2 x~.
\end{equation}
Therefore, the total number of fermions is 
\begin{equation}
\label{review:N}
    \sum\limits_{\text{ph.sp.}} u = \frac1{2\pi\hbar} \int \rmd^2 x\, u(\bx) = N~.
\end{equation}
Without loss of generality, the origin in the 1-2-plane is identified with
the center of mass of the density $u(\bx)$. 

For large $y^2+|\bx|^2$, the metric \eqref{review:metric}
approaches asymptotically AdS$_5\times S^5$, with length scale  
\begin{equation}
\label{review:L}
  L^4 = \frac1{\pi} \int \rmd^2 x\, u(\bx)~.
\end{equation}
Hence, a circular black droplet of radius $R$ corresponding to the
fermionic ground state (the Fermi sea) generates precisely 
AdS$_5 \times S^5$, with $L^2 = R$. 

From \eqref{review:N} and \eqref{review:L} one readily obtains 
\begin{equation}
\label{review:hN}
  L^4 = 2\hbar N~.
\end{equation}
Comparing this with the standard
formula, $L^4=4\pi g_s {\alpha'}^2 N$, yields the ``Planck constant'' 
\begin{equation}
\label{review:hbar}
  \hbar = 2 \pi g_s {\alpha'}^2~.
\end{equation}
The leading deviations from the asymptotic AdS$_5 \times S^5$ geometry
reveal the presence of equal angular momentum and mass \cite{Lin:2004nb}, given by
\begin{equation}
\label{review:Delta}
  \Delta=J=
  \frac1{2\pi\hbar} \int \rmd^2 x \frac{|\bx|^2}{2 \hbar} u(\bx)  
  - \frac12 \left( \frac1{2\pi\hbar} 
    \int \rmd^2 x \,u(\bx) \right)^2~.
\end{equation}
This corresponds precisely to the energy of the fermions above the
ground state. 

Finally, considering regular solutions, the quantization of 5-form
flux imposes an additional restriction on the droplet distribution. 
It can be shown that the area of any compact black or white region is
quantized in Planck units \cite{Lin:2004nb}, 
\begin{equation}
\label{review:quant}
  (\text{area}) = 2 \pi \hbar n~,
\end{equation}
with some integer $n$.

\section{Droplets, Young diagrams and limit curves}
\label{young}

In this section, we shall consider LLM solutions determined by
density functions $u(r)$ with rotational symmetry on the 1-2-plane. 
Such solutions possess two Killing vectors and are static in
the asymptotic AdS coordinates. We will show that there exists a
one-to-one map between these LLM solutions and
(piecewise) monotonic curves $X(Y)$. For regular geometries
generated by distributions of concentric black rings, these curves
precisely delimit the Young diagrams that specify the dual 1/2-BPS
microstates.\footnote{In this case, the map can be found explicitly in Fig.~1 of \cite{Ghodsi:2005ks}.} 
For general densities $u(r)$, they are the \emph{limit curves} of \cite{Balasubramanian:2005mg}.

In what follows, we use polar coordinates $(r,\phi)$ on the 1-2-plane.

\subsection{Microstates: Droplets and Young diagrams}

\begin{figure}[th]
\begin{center}%
a)
\includegraphics[bb=128 586 419 716, clip]{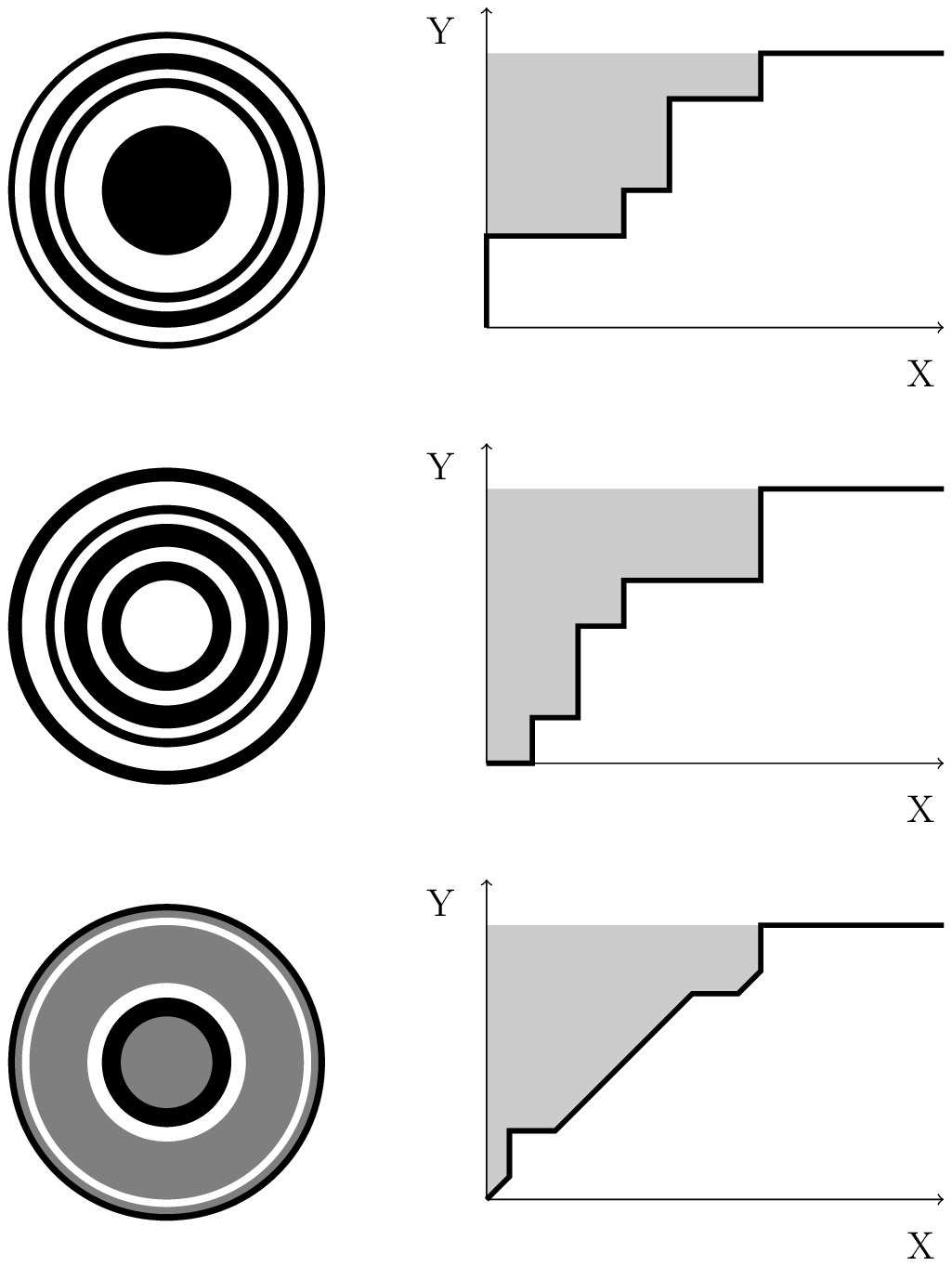}
\\[2ex]%
b)
\includegraphics[bb=128 457 419 586, clip]{Figs/rings.eps}
\end{center}
\caption{\label{young:rings}%
Examples of black ring distributions and their associated Young
diagram curves.}
\end{figure}

Let us start with a droplet configuration of $M$ concentric black rings, such
as those illustrated in Fig.~\ref{young:rings}. Since the LLM geometry
generated by such a configuration is regular, it is interpreted as a
microstate, \ie\ a pure quantum state.
The quantization of five-form flux \eqref{review:quant} introduces two
sets of integers, $b_k$ and $w_k$ ($k=1,\ldots,M$), specifying the areas
of the single black and white rings, respectively. More precisely, 
if we denote by $r_k$ and $R_k$ the
inner and outer radii of the $k$-th black ring, respectively, 
where $k=1$ stands for the innermost ring, then the integers $b_k$ and
$w_k$ are defined by
\begin{equation}
\label{young:bwdef}
  b_k = \frac1{2\pi \hbar} \pi (R_k^2 -r_k^2)~, \quad
  w_k = \frac1{2\pi \hbar} \pi (r_k^2 -R_{k-1}^2)~.
\end{equation}
Here and henceforth, we define $R_0=0$. Moreover, $r_1=0$ is an allowed value
corresponding to a configuration where the innermost ring is a disc,
as in Fig.~\ref{young:rings}a). 
The total area of all black rings specifies the total number of
fermions, so 
\begin{equation}
\label{young:sumb}
  \sum\limits_{k=1}^M b_k =N~.
\end{equation}

Using the integers $b_k$ and $w_k$, one can construct a curve that
delimits a Young diagram as follows.
Take a pen and place it at the origin of a usual $X$-$Y$
coordinate system.\footnote{We use capital letters $X$ and $Y$ in
  order to avoid confusion with the coordinates in the LLM solution.}
Without lifting the pen, draw a line $w_1$ units to the right,
followed by $b_1$ units to the top, then $w_2$ units to the right and
$b_2$ units to the top, and so on for all 
$k=1,\ldots,M$. From where the pen is then, draw a horizontal line
extending infinitely to the right. This line, which is at $Y=N$ (in as
yet unspecified units), corresponds to the infinite white plane around
the droplets. The area between the curve, the $Y$-axis and the
horizontal line at $Y=N$ (gray areas in Fig.~\ref{young:rings})
constitutes the Young diagram that specifies 
the $1/2$-BPS operator dual to the LLM
geometry. Fig.~\ref{young:rings} shows two examples of this
construction. 

In terms of the integers $b_k$ and $w_k$, the area of the Young
diagram is
\begin{equation}
\label{young:areabk}
  \Delta = \sum\limits_{k=1}^M b_k \sum\limits_{l=1}^{k} w_l~.
\end{equation}

Let us compare this formula with LLM's formula \eqref{review:Delta}.
From \eqref{review:Delta} we obtain 
\begin{align}
\notag
  \Delta &= \frac{1}{8\hbar^2} 
           \sum\limits_{k=1}^M \left( R_k^4 -r_k^4 \right) 
     - \frac{1}{8\hbar^2} \left[ 
       \sum\limits_{k=1}^M \left( R_k^2 -r_k^2 \right) \right]^2 \\
\notag
   &= \frac{1}{8\hbar^2} \sum\limits_{k=1}^M 
     \left( R_k^2 -r_k^2 \right) \left[ \left(R_k^2+r_k^2\right) 
      - \left(R_k^2-r_k^2\right)
  -2 \sum\limits_{l<k} \left(R_l^2-r_l^2\right) \right] \\  
\label{young:Delta.rings}
  &=\frac1{4\hbar^2} \sum\limits_{k=1}^M \left( R_k^2 -r_k^2 \right) 
    \sum\limits_{l=1}^k \left(r_l^2 -R_{l-1}^2\right)~,
\end{align}
which coincides with \eqref{young:areabk} after using \eqref{young:bwdef}.

\subsection{General states: Fermion densities and limit curves} 

\begin{figure}[th]
{\hfill
\includegraphics[bb=128 327 419 457, clip]{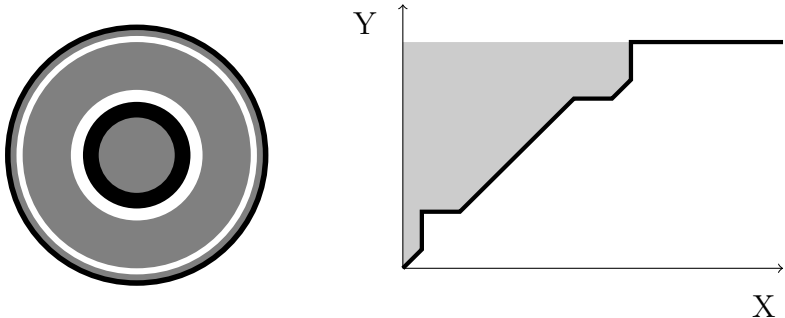}
\hfill}
\caption{\label{young:gray.rings}%
The grayscale distribution obtained by superposing the distributions
of Fig.~\ref{young:rings} with equal weights, and its associated curve.}
\end{figure}

It is possible to generalize the above construction to the case of rotationally
symmetric phase-space densities $u(r)$.
For this purpose, we follow the proposal of
\cite{Balasubramanian:2005mg} and define
two functions $Y(r)$ and $X(r)$  by\footnote{With respect to
  \cite{Balasubramanian:2005mg},
  we interchange $X$ and $Y$ and use a different normalization.} 
\begin{equation}
\label{young:xy.def}
  Y(r) = \int\limits_0^r \rmd r'\, r' u(r')~, \quad 
  X(r) = \frac12 r^2 -Y(r)~. 
\end{equation}
Notice that both $Y(r)$ and $X(r)$ are monotonically
increasing, because $0\leq u \leq 1$. The maximum value of $Y$ is 
\begin{equation}
\label{young:ymax}
  Y_\infty \equiv Y(r=\infty) = \hbar N = \frac12 L^4~.
\end{equation}
The pair $[X(r),Y(r)]$ defines a parametric curve on the $X$-$Y$
plane. It is easily seen that, for a distribution of concentric black
rings such as we considered earlier, \eqref{young:xy.def} reduces to
the construction of the Young diagram curve considered in the last
subsection, with unit length $\hbar$ on both
axes. Fig.~\ref{young:gray.rings} shows a simple example containing
areas with $u=1/2$. 

We can also go the other way around and find the fermion density
$u(r)$ associated with a given monotonic curve $Y(X)$. To do this, we
invert \eqref{young:xy.def} assuming that $X$ is a (piecewise
differentiable) function of $Y$. Then, one obtains
\begin{equation}
\label{young:u.y}
  r = \sqrt{2(X+Y)}~,\qquad u = \frac1{1+X'(Y)}~.
\end{equation}
Similarly, one can write
\begin{equation}
\label{young:u.x}
  1-u = \frac1{1+Y'(X)}~.
\end{equation}
where $Y$ is considered as a (piecewise differentiable) function of $X$.
 
Expressing the LLM energy \eqref{review:Delta} in terms of $Y$ and
$X$ using \eqref{young:xy.def} and \eqref{young:u.y}, one obtains  
\begin{equation}
\label{young:Delta.gen}
  \Delta = \frac1{\hbar^2} \int\limits_0^{Y_\infty} \rmd Y\, X(Y)~,
\end{equation}
which is, as expected, the area to the left of the curve $Y(X)$
measured in units of $\hbar$.

Now, we would like to sharpen the notion that singular geometries 
describe mixtures of microstates. To this end, we first establish that the
above construction satisfies a nice superposition property. In the
following, let $\hbar$ and $N$ (and, consequently, also $L$) be fixed. Consider a number of
rotationally symmetric density functions $u_k(r)$, each describing a
phase space distribution of $N$ fermions. These $u_k$ can correspond to microstates, but they need not.   
Let us now consider the following fermion density,
\begin{equation}
\label{young:u.superpos}
  u(r) = \sum\limits_k a_k u_k(r)~,\qquad 
  \text{with} \quad \sum\limits_k a_k =1~.
\end{equation}
Then, from \eqref{young:xy.def} follow immediately
\begin{equation}
\label{young:XY.superpos}
  X(r) = \sum\limits_k a_k X_k(r)~, \qquad  
  Y(r) = \sum\limits_k a_k Y_k(r)~, 
\end{equation}
\ie\ the superposition of fermion densities is equivalent to the
superposition of Young diagram limit curves, if the latter are regarded as
parametric curves depending on $r$. 

It is worth noting that this superposition property is consistent with
the notion of microstates as purely black and white droplet
configurations. Indeed, \eqref{young:u.superpos} implies that it is
impossible to obtain such configurations as superpositions. 

In order to show that the energy $\Delta$ associated with the mixed state with
density $u(r)$ is given by the weighted average of the energies
$\Delta_k$, one can start from \eqref{young:Delta.gen} and use
\eqref{young:xy.def} to rewrite it as\footnote{Notice that $X(Y)\neq
  \sum_k a_k X_k(Y)$.}
\begin{align}
\notag
  \Delta &= 
  \frac1{\hbar^2} \int\limits_0^{Y_\infty} \rmd Y \left( \frac12 r^2 -Y\right) 
  = \frac1{2\hbar^2} \left(  \int\limits_0^{\infty} \rmd r\,r^3 u(r)-
  Y_\infty^2 \right)\\
\label{young:Delta.superpos}
  &= \frac1{2\hbar^2} \sum\limits_k a_k \left(  \int\limits_0^{\infty}
  \rmd r\,r^3 u_k(r)- Y_\infty^2 \right) = \sum\limits_k a_k
  \Delta_k~.
\end{align}

\subsection{Entropy formula}

Mixed states possess a non-zero entropy. Therefore, it should be possible to
find a unique measure for the entropy of a mixed state, given
its density function $u(r)$. The von Neumann entropy of mixing,
$S_\text{mix} = -\sum_k a_k \ln a_k$, where one assumes that the densities $u_k$ in \eqref{young:u.superpos} 
correspond to pure states, is not a suitable definition, however. One
reason is that, given just the mixed state density $u(r)$, it is
impossible to determine unambiguously its microstate content and the
coefficients $a_k$. 

A good definition for the mixed state entropy follows from the
interpretation of the droplet plane as the phase space of
fermions. Remember that, semi-classically, the phase
space is quantized in small elements of volume $2\pi\hbar$. Let us
label these elements by an integer $n$. Then, the probabilities for finding or not finding a fermion in the phase space element $\# n$ form a Bernoulli distribution, $\{u_n, 1-u_n\}$.
Summing the von Neumann entropies of the Bernoulli distributions 
of all phase space elements, we obtain an unambiguous entropy
for the mixed state configuration. In the semi-classical limit, this
gives 
\begin{equation}
\label{young:entropy.def}
  S= - \frac{1}{2\pi\hbar} \int \rmd^2 x\, [ u\ln u +(1-u)\ln (1-u) ]~.
\end{equation}
We shall confirm in Sec.~\ref{ensembles} that \eqref{young:entropy.def}
correctly reproduces the thermodynamic entropy for ensembles of Young
diagrams.

\section{Superstar and stretched horizon}
\label{superstar}

We will consider now the construction of a stretched horizon for the
superstar geometry, which provides an estimate for the entropy of the underlying mixed state.
Let us start with the 5d reduction of the
superstar, which is a well-known extremal black hole solution of
$\mathcal{N}=2$ gauged SUGRA in five dimensions \cite{Behrndt:1998ns,Behrndt:1998jd}. 
Its metric is given by 
\begin{equation}
\label{horizon:5dmetric}
  \rmd s_5^2 = -H^{-2/3} f \rmd t^2 + H^{1/3} \left(
  f^{-1} \rmd \rho^2 +\rho^2 \rmd\Omega_3^2 \right)~,
\end{equation}
where
\begin{equation}
\label{horizon:Hf}
  H(\rho) = 1 + \omega \frac{L^2}{\rho^2}~, \qquad 
  f(\rho) = 1+ H(\rho) \frac{\rho^2}{L^2}~.
\end{equation}
For convenience, we have expressed the black hole charge by the
dimensionless $\omega$ in units of the AdS length scale, $L$. We use
$\rho$ instead of the conventional $r$ in order to avoid confusion
with the polar coordinate $r$ on the 1-2-plane in the 10d solution.

Following \cite{Suryanarayana:2004ig}, let us consider a static stretched
horizon located at constant $\rho=\rho_0$ and  wrapping the 3-sphere
in \eqref{horizon:5dmetric}. Its ``area'' is easily found to be
\begin{equation}
\label{horizon:5darea}
  \mathcal{A}_3 = 2 \pi^2 \rho_0^3 H(\rho_0)^{1/2}~.
\end{equation}
Because the 5d $\mathcal{N}=2$ gauged SUGRA can be thought of as the
compactification of 10d type-IIB SUGRA on a 5-sphere with radius $L$,
the 5d gravitational constant is 
\begin{equation}
\label{horizon:G5}
  G_5 = \frac{G_{10}}{L^5 V_{S^5}} = 
    \frac{2 \pi^4 \hbar^2}{L^5 \pi^3} =
    \frac{\pi}{2 N^2} L^3~,
\end{equation}
where we have made use of \eqref{review:hN} and \eqref{review:hbar}.
Hence, an estimate for the entropy is found as
\begin{equation}
\label{horizon:S3}
  S_\text{hor} \approx \frac{\mathcal{A}_3}{4G_5} = 
    \pi N^2 \frac{\rho_0^2}{L^2} \sqrt{\omega +\rho_0^2/L^2}~.
\end{equation}

It is instructive to repeat this calculation in ten dimensions using
the LLM form of the superstar solution. After uplifting the 5d metric
\eqref{horizon:5dmetric}, one obtains the 10d
superstar in the form \cite{Cvetic:1999xp} 
\begin{equation}
\label{horizon:10dmetric}
\begin{split}
  \rmd s_{10}^2 &= 
  -\frac1{\sqrt{D}} \left( \cos^2 \theta + D\frac{\rho^2}{L^2}
  \right)\rmd t^2  
  +\frac{2L}{\sqrt{D}} \sin^2 \theta\, \rmd t \,\rmd \phi 
  +\frac{L^2 H}{\sqrt{D}} \sin^2\theta \,\rmd \phi^2 + \\
& \quad
 + \sqrt{D} \left(f^{-1} \rmd \rho^2 +\rho^2 \rmd\Omega_3^2 \right) 
 + L^2 \sqrt{D} \,\rmd \theta^2 
 + \frac{L^2}{\sqrt{D}} \cos^2 \theta\, \rmd \tilde{\Omega}_3^2~,
\end{split}
\end{equation}
where $D = \sin^2 \theta + H\cos^2 \theta$, $\theta \in [0,\pi/2]$, and $\phi \in [0,2\pi]$. 
Upon changing coordinates to \cite{Caldarelli:2004mz}
\begin{equation}
\label{horizon:changecoords}
  y = L \rho \cos\theta~, \qquad r = L^2 \sqrt{f(\rho)} \sin\theta~,\qquad t
  \to Lt~,
\end{equation}
one finds the superstar in the standard LLM form
\eqref{review:metric}. The 1-2-plane is described in polar
coordinates ($r,\phi$). The fermion distribution generating the solution is
given by a gray disc of radius 
\begin{equation}
\label{horizon:r0}
r_0 = L^2 \sqrt{1+\omega}
\end{equation}
and uniform density 
\begin{equation}
\label{horizon:u} 
 u = (1+\omega)^{-1}~.
\end{equation}
Hence, according to \eqref{young:u.y}, the limit curve corresponding to the superstar is a straight line with slope $X'(Y)=\omega$, for $0\leq Y \leq Y_\infty$, and the delimited area is just a right angled triangle with side lengths $Y_\infty=\hbar N$ and $X(Y_\infty)=\hbar N \omega$. The energy of the superstar follows easily from  \eqref{young:Delta.gen}, 
\begin{equation}
\label{horizon:Delta}
  \Delta = \frac12 N^2 \omega~.
\end{equation}
According to our general formula \eqref{young:entropy.def}, the entropy of the underlying mixed state is 
\begin{equation}
\label{horizon:S}
  S = N [(1+\omega)\ln (1+\omega) -\omega \ln \omega]~.
\end{equation}

Let us consider now the uplift of the stretched horizon. 
It is a static 8-surface wrapping the 
two 3-spheres and forming a rotationally symmetric 2-surface above the
$y=0$ plane. Its $y$-$r$ profile follows from
\eqref{horizon:changecoords} and is the segment of an ellipse, 
\begin{equation}
\label{horizon:hordef}
  \frac{y^2}{y_H^2} + \frac{r^2}{r_H^2} = 1~,
\end{equation}
with 
\begin{equation}
\label{horizon:yrH}
  y_H = L \rho_0 ~,\qquad r_H = L^2 \sqrt{1+\omega+\rho_0^2/L^2}~.
\end{equation}

From the LLM metric \eqref{review:metric} one obtains the area of the
stretched horizon as
\begin{equation}
\label{horizon:A8}
  \mathcal{A}_8 = \int \rmd \Omega_3\, \rmd \tilde{\Omega}_3\, 
  \rmd \phi\, \rmd r\, y^2 \sqrt{1+{y'}^2} 
  \sqrt{\left(\frac14 -z^2 \right) r^2 -y^2 V_\phi^2}~,
\end{equation}
where $y' = \rmd y/\rmd r$. To continue, we still need expressions for
$z$ and $V_\phi$ on the horizon. The easiest way to find them is from the
uplift formulae \cite{Cvetic:1999xp}, which imply
\begin{align}
\label{horizon:zdef}
  z &= \frac12 \frac{\rho^2 D -L^2 \cos^2 \theta}{\rho^2 D +L^2 \cos^2
  \theta} 
  = \frac12 \frac{\rho_0^4+(\omega-1)y^2}{\rho_0^4+(\omega+1)y^2}~,\\
\label{horizon:Vdef}
  V_\phi &= -\frac{\sin^2 \theta}{\cos^2 \theta + \rho^2 D/L}
  = - \frac{L^2 \rho_0^2 -y^2}{\rho_0^4+(\omega+1)y^2}~,
\end{align}
where the second equalities in each line follow from
\eqref{horizon:changecoords}. After substituting these formulae
into \eqref{horizon:A8}, changing integration variable from $r$ to $y$
and using \eqref{horizon:hordef} and \eqref{horizon:yrH}, the
expression simplifies dramatically after some algebra. We end up with
\begin{equation}
\label{horizon:A8n}
  \mathcal{A}_8 = (2\pi^2)^2 (2\pi) \int\limits_0^{y_H} \rmd y\, 
  y^3 \frac{L^2}{\rho_0^2} \sqrt{\omega +\rho_0^2/L^2} 
  = 2 \pi^5 \rho_0^2 L^6 \sqrt{\omega +\rho_0^2/L^2}~.
\end{equation}
Hence, we obtain for the entropy
\begin{equation}
\label{horizon:S8}
  S_\text{hor} \approx \frac{\mathcal{A}_8}{4G_{10}} 
  = \frac{2 \pi^5 \rho_0^2 L^6 \sqrt{\omega +\rho_0^2/L^2}}{4(2\pi^4
  \hbar^2)} 
  = \pi N^2 \frac{\rho_0^2}{L^2} \sqrt{\omega +\rho_0^2/L^2}~,
\end{equation}
in agreement with \eqref{horizon:S3}.

As we have seen in \eqref{horizon:S}, the entropy is of order $N$. In
order to achieve this, the radius parameter $\rho_0$ must be of order
$1/\sqrt{N}$ in units of the AdS length scale $L$. More, precisely,
setting
\begin{equation}
\label{horizon:choose}
  \frac{\rho_0^2}{L^2} = \frac{c(\omega)}{\pi N}~,
\end{equation}
where $c(\omega)$ can depend on the slope of the limit curve, and neglecting the second term in the square
root (we have large $N$), we obtain
\begin{equation}
\label{horizon:Sfinal}
  S_\text{hor} \approx c(\omega) N\sqrt{\omega} = c(\omega) \sqrt{2\Delta}~,
\end{equation}
where $\Delta$ is the superstar energy \eqref{horizon:Delta}.

One can justify the form \eqref{horizon:choose} by considering the minimum fluctuation of the number of columns of a Young diagram \cite{Suryanarayana:2004ig}, which is, obviously, $\delta n_c=1$. Fluctuations of the number of columns translate into fluctuations at the upper end of the limit curve triangle. Moreover, as the upper end of the limit curve corresponds to the border of the gray disc of non-zero fermion density, we can interpret these fluctuations as fluctuations of the disc's radius. In particular, let us use \eqref{young:xy.def} to write\footnote{$Y_\infty$ is fixed.}
\begin{equation}
\label{ens:delta.r}
  \delta r_0^2 = 2\, \delta X(Y_\infty) = 2 \hbar\, \delta n_c~.
\end{equation}
Considering a stretched horizon of the form \eqref{horizon:hordef}, we identify 
\begin{equation}
\label{ens:rH.def}
  r_H^2 = r_0^2 + \delta r_0^2~.
\end{equation}
Thus, using also \eqref{horizon:r0} and \eqref{horizon:yrH}, we determine the parameter $\rho_0$ as 
\begin{equation}
\label{ens:rho0def}
  \frac{\rho_0^2}{L^2} = \frac1{L^4} \left( r_H^2 -r_0^2 \right) = \frac1{L^4} \delta r_0^2 =
  \frac1N \delta n_c~, 
\end{equation}
implying $c(\omega)=\pi$ in \eqref{horizon:choose} and \eqref{horizon:Sfinal}. 

Using the minimum fluctuation of a statistical ensemble is, however, not very reasonable.  
In Sec.~\ref{ensembles}, we will check whether the mean size of fluctuations in the superstar ensemble is in agremment with \eqref{horizon:choose}.  

\section{Ensembles of Young diagrams}
\label{ensembles}

In this section, we shall consider some statistical (thermodynamic) ensembles of Young diagrams. Our main interest is in ensembles, whose \emph{limit curves} are in agreement with the superstar solution. We will, however, start by considering the grand canonical ensemble of \cite{Balasubramanian:2005mg} and verify that the fermionic entropy \eqref{young:entropy.def} reproduces the thermodynamic entropy of the ensemble. 

First, let us state how we describe Young diagrams. The most convenient way for our purposes to describe a given Young diagram is to specify $N$ numbers $c_j$, $j=1,2,\ldots,N$, which count the number of columns of length $j$. The maximum possible length of any column is $N$. The total number of columns of the diagram thus specified is 
\begin{equation}
\label{ens:nc}
  n_c = \sum\limits_{j=1}^N c_j~, 
\end{equation}
whereas the total number of boxes of the diagram is
\begin{equation}
\label{ens:nboxes}
  E = \sum\limits_{j=1}^N  j\, c_j~. 
\end{equation}
For $1/2$-BPS states in $\mathcal{N}=4$ SYM with gauge group $U(N)$, the $c_j$'s can take any non-negative integer value, $c_j=0,1,2,\ldots$. For gauge group $SU(N)$, the restriction $\tr Z=0$\footnote{$Z$ is the   complex combination of two of the six real scalar fields, \eg $Z=X^1+iX^4$, which transform in the adjoint of the gauge group.}
removes all diagrams with $c_1>0$. In what follows, we shall consider $U(N)$. In the large-$N$ (thermodynamic) limit, the differences between the $U(N)$ and $SU(N)$ results are subleading in $1/N$.

In the thermodynamic limit, we replace the integers $j$ and $c_j$ by the continuous variables 
of the limit curve, such that
\begin{equation}
\label{ens:XYdef} 
   \hbar j \to  Y~, \qquad \hbar \sum\limits_{i=1}^j c_i \to X(Y) 
   \quad \Rightarrow \quad c_j \to X'(Y)~,
\end{equation}
and take the limit such that
\begin{equation}
\label{ens:thermlim}
  \hbar \to 0~,\qquad N\to \infty~,\qquad \text{with $\hbar N=Y_\infty$ fixed.}
\end{equation}
The last replacement in \eqref{ens:XYdef} comes very handy in view of the relation \eqref{young:u.y}.

\subsection{Grand canonical ensemble} 
\label{end:grand}

Balasubramanian \emph{et al.}\ \cite{Balasubramanian:2005mg} studied a grand canonical ensemble, in which the mean energy, $\Delta=\vev{E}$, and the mean number of colums, $N_c=\vev{n_c}$, were fixed using Lagrange multipliers. They found that the superstar limit curve is obtained for infinite ``temperature''. In the following, we shall briefly review their calculations for arbitrary temperature, in order to verify our proposal for the entropy \eqref{young:entropy.def}, and specialize to infinite temperature in Sec.~\ref{ens:grand.superstar}.

Let us start with the grand canonical partition function\footnote{It is useful for the thermodynamic limit to 
introduce as Lagrange multiplier $\alpha=\beta N$, where $\beta$ is the inverse ``temperature''.}
\begin{equation}
\label{ens:Zgrand}
  Z = \sum\limits_{\{c_j\}} \e{-\alpha E/N -\lambda n_c} = 
    \sum\limits_{\{c_j\}} \e{-\sum\limits_{j=1}^N (\alpha j/N +\lambda) c_j}~.
\end{equation}
After performing the sum over the $c_j$'s in \eqref{ens:Zgrand}, we obtain\footnote{This holds for gauge group $U(N)$. For $SU(N)$, the product would start with $j=2$.}
\begin{equation}
\label{ens:Zgr2}
  Z = \prod\limits_{j=1}^N  \left[1-\e{-(\alpha j/N+\lambda)}\right]^{-1}~.
\end{equation}
To continue, we consider $\ln Z$ and take the thermodynamic limit \eqref{ens:thermlim}, with the result 
\begin{equation}
\label{ens:Zgr3}
  \ln Z = \frac{N}{\alpha} \left[ \rmLi_2(\e{-\lambda}) - \rmLi_2(\e{-(\alpha+\lambda)}) \right]~,
\end{equation}
where 
\begin{equation}
\label{ens:Li}
  \rmLi_2(x) = \int\limits_1^{1-x} \rmd t \frac{\ln t}{1-t} = \sum\limits_{n=1}^\infty \frac{x^n}{n^2}
\end{equation}
is the dilogarithm function.

The Lagrange multipliers are fixed imposing the mean values
\begin{equation}
\label{ens:Lagrange1}
  \frac{\Delta}{N} =\frac{\vev{E}}{N} = -\frac{\partial}{\partial \alpha} \ln Z 
    = \frac1{\alpha} \ln Z +\frac{N}{\alpha} \ln \left(1-\e{-(\alpha+\lambda)}\right)
\end{equation}
and 
\begin{equation}
\label{ens:Lagrange2} 
  N_c = \vev{n_c} = -\frac{\partial}{\partial \lambda} \ln Z
      = \frac{N}{\alpha} \ln \frac{1-\e{-(\alpha+\lambda)}}{1-\e{-\lambda}}~.
\end{equation}
Then, the thermodynamic entropy is found to be
\begin{align}
\notag
  S_{\text{therm}} &= \ln Z +\alpha \frac{\Delta}N +\lambda N_c \\
\label{ens:Stherm}
  &= \frac{N}{\alpha} \left[ 2 \rmLi_2(\e{-\lambda}) - 2 \rmLi_2(\e{-(\alpha+\lambda)}) 
  +(\alpha+\lambda) \ln( 1-\e{-(\alpha+\lambda)} ) - \lambda \ln( 1-\e{-\lambda} ) \right]~.
\end{align}

We would like to verify that the semi-classical fermionic entropy \eqref{young:entropy.def} correctly reproduces the thermodynamic entropy \eqref{ens:Stherm}. Taking into account \eqref{young:u.y} and \eqref{ens:XYdef}, we need to find the mean value $\vev{c_j}$,
\begin{equation}
\label{ens:vev.cj}
  \vev{c_j} = \frac{1}{Z} \sum\limits_{\{c_i\}} c_j \e{-\sum\limits_{i=1}^N (\alpha i/N +\lambda) c_i} 
  = \left(\e{\alpha j/N +\lambda} -1 \right)^{-1}~.   
\end{equation}
Thus, the fermion density \eqref{young:u.y} is\footnote{For any smooth function $f(x)$, we have $\vev{f(x)} = f(\vev{x}) +\cdots$, where the ellipses denote terms that vanish in the thermodynamic limit.}
\begin{equation}
\label{ens:u}
  u = 1- \e{-(\alpha j/N+\lambda)}~,
\end{equation}
and the entropy \eqref{young:entropy.def} becomes, after making the replacement \eqref{ens:XYdef},
\begin{align}
\notag
  S &= -\frac{1}{\hbar} \int\limits_0^{Y_\infty} \rmd Y 
  \left[ \ln u +\frac{1-u}{u} \ln (1-u) \right] \\
\label{ens:Sferm}
  &= \frac{N}{\alpha} \left[ \rmLi_2(\e{-\lambda}) - \rmLi_2(\e{-(\alpha+\lambda)}) - 
  \rmLi_2(1-\e{-\lambda}) + \rmLi_2(1-\e{-(\alpha+\lambda)}) \right]~.
\end{align}
Eq.~\eqref{ens:Sferm} is found to agree with \eqref{ens:Stherm} by means of the identity \cite{Abramowitz}
\begin{equation}
\label{ens:Li2.ident}
  \rmLi_2(x) + \rmLi_2(1-x) = -\ln(1-x)\ln x +\frac{\pi^2}6~.
\end{equation}

\subsection{Grand canonical superstar ensemble}
\label{ens:grand.superstar}

The superstar geometry corresponds to a triangular limit curve with constant slope $X'(Y)=\vev{c_j}$. From \eqref{ens:vev.cj} we see that this implies $\alpha=0$. For this value of $\alpha$, the expressions for the thermodynamic quantities simplify. Defining 
\begin{equation}
\label{ens:omega:def}
  \omega = (\e{\lambda}-1)^{-1}~,
\end{equation}
the mean energy \eqref{ens:Lagrange1} and number of columns \eqref{ens:Lagrange2} become simply
\begin{equation}
\label{ens:Delta.nc.sup}
  \Delta = \frac{\omega}2 N^2~, \qquad
  N_c = \omega N~,
\end{equation}
and the entropy \eqref{ens:Sferm} reduces to
\begin{equation}
\label{ens:entropy}
\begin{split}
  S &= N\left[ (1+\omega) \ln (1+\omega) -\omega \ln \omega \right] \\
  &= \sqrt{2\Delta} \left[ (\omega^{1/2} +\omega^{-1/2}) \ln (1+\omega) -\omega^{1/2} \ln \omega \right]~.
\end{split}
\end{equation}
It is interesting to note that, for fixed energy $\Delta$, the entropy is invariant under $\omega\to 1/\omega$,  exchanging rows with columns. The energy and entropy are, as expected, in agreement with the superstar formulae \eqref{horizon:Delta} and \eqref{horizon:S}.

So far, we have dealt only with the mean quantities, which are independent of the choice of ensemble (canonical, grand canonical or micro canonical). Differences between the ensembles appear when fluctuations around the mean values are considered, which are suppressed in the thermodynamic limit. However, the size of fluctuations is of interest for the construction of a stretched horizon around the singularity of the dual geometry. This is because the stretched horizon roughly surrounds the space-time region, where the underlying microstates significantly differ from each other, and where, therefore, the singular solution becomes unreliable. 

Let us, for example, consider the variance of the total number of columns,
\begin{equation}
\label{ens:var.nc}
  \Var n_c = \vev{n_c^2} - \vev{n_c}^2 
  = \sum\limits_{i,j=1}^N \left( \vev{c_i c_j} - \vev{c_i} \vev{c_j} \right)~.
\end{equation}
It is straightforward to find the correlations
\begin{equation}
\label{ens:cj.corr}
  \vev{c_i c_j} -\vev{c_i} \vev{c_j} = 
  \begin{cases} 
     \omega(1+\omega) \quad & \text{for $i=j$,}\\
     0 & \text{otherwise.}
  \end{cases}
\end{equation}
Hence, we obtain
\begin{equation}
\label{ens:var.nc2}
  \Var n_c = N \omega (1+\omega)~.
\end{equation}

Let us now estimate the size of a suitable stretched horizon. Setting $\delta n_c = \sqrt{\Var n_c}$ and using \eqref{ens:rho0def}, we obtain 
\begin{equation}
\label{ens:rho0}
  \frac{\rho_0^2}{L^2} = \frac{\sqrt{\omega(1+\omega)}}{\sqrt{N}}~.
\end{equation}
This is too large in the large-$N$ limit, because it would imply that the entropy \eqref{horizon:S8} associated with the stretched horizon is of order $N^{3/2}$, whereas \eqref{ens:entropy} is of order $N$. Therefore, we conclude that the grand canonical ensemble at infinite temperature contains far too many microstates.  In the following section, we find a way to improve on this point.

\subsection{Restricted superstar ensemble}
\label{ens:res.superstar}

In this section, we introduce an ensemble that exhibits the mean values of the
grand canonical superstar ensemble, while containing smaller fluctuations. We
shall call it henceforth the \emph{restricted superstar ensemble}.
The main idea is to incorporate the condition on the number of
columns directly into the sum over configurations instead of imposing
it via a Lagrange multiplier. Hence, let us define the ensemble\footnote{For gauge group $SU(N)$, 
  the additional constraint $c_1=0$ has to be applied.}
\begin{equation}
\label{micro:ensemble}
  \{ c_j \}^* = \left\{ c_j=0,1,2,\ldots ;
   \sum\limits_{j=1}^N c_j \leq N_c \right\}~.
\end{equation}
In words, the restricted superstar ensemble contains all configurations
whose total number of columns does not exceed $N_c$. The choice of
``does not exceed'' as opposed to ``is equal to'' puts $N_c$ on the same footing for the columns 
as $N$ for the rows and renders the
ensemble definition symmetric under $N \leftrightarrow
N_c$.\footnote{Remember that Young diagrams with $c_N=0$ are allowed.} 

The mean energy is imposed in the canonical way. This means that, in order to obtain the superstar limit curve, the restricted superstar ensemble should be considered at infinite temperature, so we set
$\alpha=0$ from the start. This is a fortunate circumstance, because
expectation values can be computed quite easily. Let us start with the
partition function 
\begin{equation}
\label{micro:Zdef}
  Z = \sum\limits_{\{c_j\}^*} 1 = 
      \sum\limits_{c_N=0}^{N_c} \sum\limits_{c_{N-1}=0}^{N_c-c_N} 
      \cdots \sum\limits_{c_1=0}^{N_c-c_N-\cdots -c_2} 1~.
\end{equation}
The sums can be performed by making use of the formula
\cite{Gradshteyn}
\begin{equation}
\label{micro:binomsum}
  \sum\limits_{k=0}^m \binom{n+k}{k} = \binom{n+m+1}{n+1}~,
\end{equation}
with the final result
\begin{equation}
\label{micro:Z}
  Z = \binom{N+N_c}{N}~.
\end{equation}

In order to calculate expectation values of products of the
$c_j$'s, in addition to \eqref{micro:binomsum}, one also needs the
identity
\begin{equation}
\label{micro:binomid}
  (a+1) \binom{a}{b} = (b+1) \binom{a+1}{b+1}~.
\end{equation}
The results for the one- and two-point correlators are 
\begin{align}
\label{micro:av_j}
  \vev{c_j} &= \frac{N_c}{N+1}~, \\
\label{micro:av_jj}
  \vev{c_j^2} &= \frac{N_c(N+2N_c)}{(N+1)(N+2)}~,\\
\label{micro:av_ij}
  \vev{c_i c_j} &= \frac{N_c(N_c-1)}{(N+1)(N+2)} 
  \qquad \text{for $i\neq j$}~.
\end{align}
We note that, in contrast to the grand canonical superstar ensemble, the
numbers of columns of different lengths are now correlated due to the
ensemble restriction \eqref{micro:ensemble}.

The mean energy, obtained from \eqref{ens:nboxes} and \eqref{micro:av_j}, reads
\begin{equation}
\label{micro:delta}
  \Delta = \frac12 N N_c~.
\end{equation}
This is as expected and agrees with the grand canonical superstar ensemble
although, now, there was no need to take the large-$N$ limit. To compare the
entropy, we consider the thermodynamic limit, keeping the slope
\begin{equation}
\label{micro:omega}
  \omega = \frac{N_c}{N+1} 
\end{equation}
fixed. We find
\begin{equation}
\label{micro:S}
  S= \ln Z \approx N \ln (1+\omega) + N_c \ln (1+\omega^{-1})~,
\end{equation}
where only the leading term in $N$ has been written. This agrees
with \eqref{ens:entropy}, as expected.

We want to verify now whether the estimated size of the stretched horizon determined by the magnitude of fluctuations is in agreement with the thermodynamic entropy. Therefore,
let us consider again the fluctuations of the number of
columns. The average number of columns is found to be 
\begin{equation}
\label{micro:nc}
  \vev{n_c} = \frac{N N_c}{N+1}=\omega N~,
\end{equation}
with a variance
\begin{equation}
\label{micro:varnc}
  \Var n_c = \frac{N N_c (N+N_c+1)}{(N+1)^2(N+2)} 
  = \frac{N}{N+2} \omega(1+\omega) \approx \omega(1+\omega).
\end{equation}
where the last equality holds in the large-$N$ limit. Translating this into an estimate for the size of the stretched horizon by means of \eqref{ens:rho0def}, we find
\begin{equation}
\label{micro:rho0.res}
  \frac{\rho_0^2}{L^2} = \frac1N \sqrt{\Var n_c} = \frac{\sqrt{\omega(1+\omega)}}{N}~.
\end{equation}
This is of the correct order of magnitude in $N$ and implies an estimated horizon entropy \eqref{horizon:S8}
\begin{equation}
\label{micro:Shor}
  S_\text{hor} \approx N \pi \omega \sqrt{1+\omega} = \pi \sqrt{2\Delta} \sqrt{\omega(1+\omega)}~.
\end{equation}

A slightly smaller estimate can be obtained by realizing that the restricted ensemble \eqref{micro:ensemble} does not contain Young diagrams with more than $N_c$ columns. That is, in order to enclose all mircostate Young diagrams of the restricted ensemble, we can safely place the stretched horizon at [c.f.\ \eqref{young:xy.def}]
\begin{equation}
\label{micro:rH.new}
  r_H^2 = 2 ( \hbar N_c + \hbar N) = L^4 \left( \omega +\frac{\omega}N +1 \right )~,
\end{equation}
whereas the radius of the gray disc is identified with
\begin{equation} 
\label{micro:r0.new}
  r_0^2 = 2 (\hbar \vev{n_c} +\hbar N) = L^4 ( \omega +1 )~. 
\end{equation}
Thus, we obtain 
\begin{equation}
\label{micro:rho0.new}
  \frac{\rho_0^2}{L^2} = \frac1{L^4} \left(r_H^2 -r_0^2 \right)= \frac{\omega}{N}~,
\end{equation}
and
\begin{equation}
\label{micro:Shor.new}
  S_\text{hor} \approx N \pi \omega^{3/2} = \pi \omega \sqrt{2\Delta}~.
\end{equation}

\section{Conclusions and Outlook}
\label{conclusions}

In this paper, we have considered pure and mixed state configurations of the LLM class of solutions describing the $1/2$-BPS sector of $\mathcal{N}=4$ super Yang Mills theory. Pure state geometries are generated by distributions of concentric rings of droplets, while mixed states are characterized by some density function $u(r)$, which is identified with the semi-classical one-fermion phase space density. We have constructed a one-to-one correspondence between these density functions and (piecewise) monotonic curves that, in the case of pure states, precisely delimit the corresponding Young diagrams. Inspired by the fermionic phase space picture, we have proposed a unique entropy formula, eq.~\eqref{young:entropy.def}, and verified that it reproduces the thermodynamic entropy of a general grand canonical ensemble of Young diagrams. 

Further, we have studied the mean size of fluctuations in thermodynamic ensembles of Young diagrams in order to produce an estimate for the size of a stretched horizon that encloses the fluctuations of the microstate geometries. Our results show that the grand canonical superstar ensemble of \cite{Balasubramanian:2005mg} has far too many fluctuations implying that the entropy would grow like $N^{3/2}$ with the number of fermions, while the thermodynamic formula yields a growth like $N$. To improve upon this, we have introduced a restricted superstar ensemble by imposing an upper limit on the number of columns. For this ensemble, the estimated entropy scales correctly with $N$ and differs from the thermodynamic entropy by a function of the black hole charge parameter $\omega$. As a drawback, we observe that our estimates \eqref{micro:Shor} and \eqref{micro:Shor.new} are not invariant under an exchange of rows and columns, keeping the energy fixed. It is conceivable that using the micro canonical ensemble (with the restriction $n_c \leq N_c$) would improve on this point, but that ensemble is too difficult to handle explicitly. Furthermore, as the stretched horizon can only yield an order-of-magnitude estimate, we would not expect it to reproduce the entropy entirely. 

The superstar belongs to the class of extremal small black holes discussed in the introduction. Including higher derivative corrections to SUGRA, it is expected that a horizon with finite area is generated. It would be very interesting to obtain a geometric entropy, which can be compared with the thermodynamic entropy of the superstar ensemble. To do this, one could, for example, consider the recently constructed higher derivative corrections to five-dimensional SUGRA \cite{Hanaki:2006pj}. Using Sen's entropy function approach in conformal supergravity language, similar to \cite{Castro:2007sd}, seems very promising. Another very interesting question is whether regular solutions continue to be generated by a distribution of droplets, once higher derivative corrections are included.

\section*{Acknowledgments}
It is a pleasure to thank D.~Klemm, G.~Maiella and P.~Silva for interesting discussions. 
This work was supported in part by the European Community's Human Potential Programme under Contract MRTN-CT-2004-005104 `Constituents, fundamental forces and symmetries of the universe', and by the Italian Ministry of Education and Research (MIUR), project 2005-023102.

%\bibliographystyle{JHEP}
%\bibliography{LLM}

\begin{thebibliography}{10}

\bibitem{Mathur:2005ai}
S.~D. Mathur, {\it The quantum structure of black holes},  {\em Class. Quant.
  Grav.} {\bf 23} (2006) R115,
  [\href{http://xxx.lanl.gov/abs/hep-th/0510180}{{\tt hep-th/0510180}}].

\bibitem{Dabholkar:2004dq}
A.~Dabholkar, R.~Kallosh, and A.~Maloney, {\it A stringy cloak for a classical
  singularity},  {\em JHEP} {\bf 12} (2004) 059,
  [\href{http://xxx.lanl.gov/abs/hep-th/0410076}{{\tt hep-th/0410076}}].

\bibitem{Sen:2005wa}
A.~Sen, {\it Black hole entropy function and the attractor mechanism in higher
  derivative gravity},  {\em JHEP} {\bf 09} (2005) 038,
  [\href{http://xxx.lanl.gov/abs/hep-th/0506177}{{\tt hep-th/0506177}}].

\bibitem{Morales:2006gm}
J.~F. Morales and H.~Samtleben, {\it Entropy function and attractors for ads
  black holes},  {\em JHEP} {\bf 10} (2006) 074,
  [\href{http://xxx.lanl.gov/abs/hep-th/0608044}{{\tt hep-th/0608044}}].

\bibitem{Sinha:2006yy}
A.~Sinha and N.~V. Suryanarayana, {\it Extremal single-charge small black
  holes: Entropy function analysis},  {\em Class. Quant. Grav.} {\bf 23} (2006)
  3305--3322, [\href{http://xxx.lanl.gov/abs/hep-th/0601183}{{\tt
  hep-th/0601183}}].

\bibitem{Kraus:2005vz}
P.~Kraus and F.~Larsen, {\it Microscopic black hole entropy in theories with
  higher derivatives},  {\em JHEP} {\bf 09} (2005) 034,
  [\href{http://xxx.lanl.gov/abs/hep-th/0506176}{{\tt hep-th/0506176}}].

\bibitem{Castro:2007sd}
A.~Castro, J.~L. Davis, P.~Kraus, and F.~Larsen, {\it 5d attractors with higher
  derivatives},  \href{http://xxx.lanl.gov/abs/hep-th/0702072}{{\tt
  hep-th/0702072}}.

\bibitem{Astefanesei:2006dd}
D.~Astefanesei, K.~Goldstein, R.~P. Jena, A.~Sen, and S.~P. Trivedi, {\it
  Rotating attractors},  {\em JHEP} {\bf 10} (2006) 058,
  [\href{http://xxx.lanl.gov/abs/hep-th/0606244}{{\tt hep-th/0606244}}].

\bibitem{Mathur:2005zp}
S.~D. Mathur, {\it The fuzzball proposal for black holes: An elementary
  review},  {\em Fortsch. Phys.} {\bf 53} (2005) 793--827,
  [\href{http://xxx.lanl.gov/abs/hep-th/0502050}{{\tt hep-th/0502050}}].

\bibitem{Balasubramanian:2005mg}
V.~Balasubramanian, J.~de~Boer, V.~Jejjala, and J.~Simon, {\it The library of
  babel: On the origin of gravitational thermodynamics},  {\em JHEP} {\bf 12}
  (2005) 006, [\href{http://xxx.lanl.gov/abs/hep-th/0508023}{{\tt
  hep-th/0508023}}].

\bibitem{Lin:2004nb}
H.~Lin, O.~Lunin, and J.~M. Maldacena, {\it Bubbling ads space and 1/2 bps
  geometries},  {\em JHEP} {\bf 10} (2004) 025,
  [\href{http://xxx.lanl.gov/abs/hep-th/0409174}{{\tt hep-th/0409174}}].

\bibitem{Corley:2001zk}
S.~Corley, A.~Jevicki, and S.~Ramgoolam, {\it Exact correlators of giant
  gravitons from dual n = 4 sym theory},  {\em Adv. Theor. Math. Phys.} {\bf 5}
  (2002) 809--839, [\href{http://xxx.lanl.gov/abs/hep-th/0111222}{{\tt
  hep-th/0111222}}].

\bibitem{Berenstein:2004kk}
D.~Berenstein, {\it A toy model for the ads/cft correspondence},  {\em JHEP}
  {\bf 07} (2004) 018, [\href{http://xxx.lanl.gov/abs/hep-th/0403110}{{\tt
  hep-th/0403110}}].

\bibitem{Caldarelli:2004mz}
M.~M. Caldarelli, D.~Klemm, and P.~J. Silva, {\it Chronology protection in
  anti-de sitter},  {\em Class. Quant. Grav.} {\bf 22} (2005) 3461--3466,
  [\href{http://xxx.lanl.gov/abs/hep-th/0411203}{{\tt hep-th/0411203}}].

\bibitem{Milanesi:2005tp}
G.~Milanesi and M.~O'Loughlin, {\it Singularities and closed time-like curves
  in type iib 1/2 bps geometries},  {\em JHEP} {\bf 09} (2005) 008,
  [\href{http://xxx.lanl.gov/abs/hep-th/0507056}{{\tt hep-th/0507056}}].

\bibitem{Buchel:2004mc}
A.~Buchel, {\it Coarse-graining 1/2 bps geometries of type iib supergravity},
  {\em Int. J. Mod. Phys.} {\bf A21} (2006) 3495--3502,
  [\href{http://xxx.lanl.gov/abs/hep-th/0409271}{{\tt hep-th/0409271}}].

\bibitem{Suryanarayana:2004ig}
N.~V. Suryanarayana, {\it Half-bps giants, free fermions and microstates of
  superstars},  {\em JHEP} {\bf 01} (2006) 082,
  [\href{http://xxx.lanl.gov/abs/hep-th/0411145}{{\tt hep-th/0411145}}].

\bibitem{Silva:2005fa}
P.~J. Silva, {\it Rational foundation of gr in terms of statistical mechanic in
  the ads/cft framework},  {\em JHEP} {\bf 11} (2005) 012,
  [\href{http://xxx.lanl.gov/abs/hep-th/0508081}{{\tt hep-th/0508081}}].

\bibitem{Ghodsi:2005ks}
A.~Ghodsi, A.~E. Mosaffa, O.~Saremi, and M.~M. Sheikh-Jabbari, {\it Lll vs.
  llm: Half bps sector of n = 4 sym equals to quantum hall system},  {\em Nucl.
  Phys.} {\bf B729} (2005) 467--491,
  [\href{http://xxx.lanl.gov/abs/hep-th/0505129}{{\tt hep-th/0505129}}].

\bibitem{Takayama:2005bc}
Y.~Takayama and K.~Yoshida, {\it Bubbling 1/2 bps geometries and penrose
  limits},  {\em Phys. Rev.} {\bf D72} (2005) 066014,
  [\href{http://xxx.lanl.gov/abs/hep-th/0503057}{{\tt hep-th/0503057}}].

\bibitem{Takayama:2005yq}
Y.~Takayama and A.~Tsuchiya, {\it Complex matrix model and fermion phase space
  for bubbling ads geometries},  {\em JHEP} {\bf 10} (2005) 004,
  [\href{http://xxx.lanl.gov/abs/hep-th/0507070}{{\tt hep-th/0507070}}].

\bibitem{Maoz:2005nk}
L.~Maoz and V.~S. Rychkov, {\it Geometry quantization from supergravity: The
  case of 'bubbling ads'},  {\em JHEP} {\bf 08} (2005) 096,
  [\href{http://xxx.lanl.gov/abs/hep-th/0508059}{{\tt hep-th/0508059}}].

\bibitem{Giombi:2005zq}
S.~Giombi, M.~Kulaxizi, R.~Ricci, and D.~Trancanelli, {\it Half-bps geometries
  and thermodynamics of free fermions},  {\em JHEP} {\bf 01} (2007) 067,
  [\href{http://xxx.lanl.gov/abs/hep-th/0512101}{{\tt hep-th/0512101}}].

\bibitem{Liu:2006pd}
J.~T. Liu and W.~Sabra, {\it All 1/2 bps solutions of iib supergravity with
  so(4) x so(4) isometry},  \href{http://xxx.lanl.gov/abs/hep-th/0611244}{{\tt
  hep-th/0611244}}.

\bibitem{Sato:2007zu}
M.~Sato, {\it On-shell action for bubbling geometries},
  \href{http://xxx.lanl.gov/abs/hep-th/0703063}{{\tt hep-th/0703063}}.

\bibitem{Chen:2007gh}
H.-Y. Chen, D.~H. Correa, and G.~A. Silva, {\it Geometry and topology of bubble
  solutions from gauge theory},
  \href{http://xxx.lanl.gov/abs/hep-th/0703068}{{\tt hep-th/0703068}}.

\bibitem{Brown:2007bb}
T.~W. Brown, {\it Half-bps su(n) correlators in n=4 sym},
  \href{http://xxx.lanl.gov/abs/hep-th/0703202}{{\tt hep-th/0703202}}.

\bibitem{Behrndt:1998ns}
K.~Behrndt, A.~H. Chamseddine, and W.~A. Sabra, {\it Bps black holes in n = 2
  five dimensional ads supergravity},  {\em Phys. Lett.} {\bf B442} (1998)
  97--101, [\href{http://xxx.lanl.gov/abs/hep-th/9807187}{{\tt
  hep-th/9807187}}].

\bibitem{Behrndt:1998jd}
K.~Behrndt, M.~Cvetic, and W.~A. Sabra, {\it Non-extreme black holes of five
  dimensional n = 2 ads supergravity},  {\em Nucl. Phys.} {\bf B553} (1999)
  317--332, [\href{http://xxx.lanl.gov/abs/hep-th/9810227}{{\tt
  hep-th/9810227}}].

\bibitem{Cvetic:1999xp}
M.~Cvetic {\em et~al.}, {\it Embedding ads black holes in ten and eleven
  dimensions},  {\em Nucl. Phys.} {\bf B558} (1999) 96--126,
  [\href{http://xxx.lanl.gov/abs/hep-th/9903214}{{\tt hep-th/9903214}}].

\bibitem{Abramowitz}
M.~Abramowitz and I.~A. Stegun, eds., {\em Handbook of Mathematical Functions}.
\newblock Dower Publ., New York.

\bibitem{Gradshteyn}
I.~S. Gradshteyn and I.~M. Ryzhik, {\em Table of Integrals, Series and
  Products}.
\newblock Academic Press, New York, 5~ed.

\bibitem{Hanaki:2006pj}
K.~Hanaki, K.~Ohashi, and Y.~Tachikawa, {\it Supersymmetric completion of an
  r**2 term in five- dimensional supergravity},
  \href{http://xxx.lanl.gov/abs/hep-th/0611329}{{\tt hep-th/0611329}}.

\end{thebibliography}
\providecommand{\href}[2]{#2}\begingroup\raggedright\endgroup

\end{document}